\documentclass[aps,prb,twocolumn,amsmath,amssymb]{revtex4}
\usepackage{amssymb,amsmath}
\usepackage{bm,graphicx}
\usepackage{natbib}
\usepackage{mathptmx}
\usepackage{pxfonts}
\usepackage{times}
\usepackage{fontenc}
\usepackage{graphicx}
\usepackage{Jonasmacros}

\usepackage{Jonasmacros}

\usepackage[colorlinks,citecolor=blue,linkcolor=Fuchsia]{hyperref}
\usepackage[usenames,dvipsnames,svgnames]{xcolor}

\begin{document}
\date{\today}
\title{Electron Paramagnetic Resonance of Single Magnetic Moment on a Surface}

\author{P. Berggren}
\affiliation{Department of Physics and Astronomy, Uppsala University, Box 516, SE-751 21\ \ Uppsala}

\author{J. Fransson}
\email{Jonas.Fransson@fysik.uu.se}
\affiliation{Department of Physics and Astronomy, Uppsala University, Box 516, SE-751 21\ \ Uppsala}

\begin{abstract}
We address electron spin resonance of single magnetic moments in a tunnel junction using time-dependent electric fields and spin-polarized current. We show that the tunneling current directly depends on the local magnetic moment and that the frequency of the external electric field mixes with the characteristic Larmor frequency of the local spin. The importance of the spin-polarized current induced anisotropy fields acting on the local spin moment is, moreover, demonstrated. Our proposed model thus explains the absence of an electron spin resonance for a half integer spin, in contrast with the strong signal observed for an integer spin.
\end{abstract}
\maketitle

\section*{INTRODUCTION}
Unambiguous and direct measurements of single spin moments remains an elusive goal which has yet to be reproducibly demonstrated. Upon approaching the quantum limit for  magnetic entities and magnetic interactions, the ability to make distinct determinations of single magnetic moments is crucial to deeper understand the magnetic environment.

In 1989, Manassen et al \cite{manassen1989} measured current-current correlations induced peaks in the power spectra of the tunneling current associated with precession of a local paramagnetic moment (electron spin resonance -- ESR), using scanning tunneling microscopy (STM), ESR-STM. While controversial at the time, these measurements have not only been refined \cite{manassen1997,messina2007,mannini2007,komeda2008,manassen2014} and put into theoretical context \cite{QIP.1.355,PhilMagB.82.1291,PhysRevB.66.195416}, but also been independently reproduced in different systems \cite{durkan2002,durkan2004,manniniLang2007,krukowski2010,biointerfaces.10.031001}. For a more thorough review we refer to Ref. \onlinecite{balatsky2012}.

Read-out of a single paramagnetic moment has been achieved in different all-electrical designs, e.g., semi-conducting field effect transistors \cite{xiao2004}, double quantum dots working in the Pauli spin-blockade regime \cite{koppens2006,pioro2008} and spin-valley regime \cite{hao2014}, as well as in optical measurements of, e.g., nitrogen-vacancy centers in diamond \cite{epstein2005}. Theoretically, the field has witnessed a huge progress for various single spin set-ups \cite{molotkov1992,prioli1995,engel2001,mozyrsky2002,zhu2002,levitov2003,bulaevskii2003,PhysRevLett.90.040401,manassen2004,balatsky2006,golub2013}. The full potential of single spin ESR, especially all electrical, has yet to be considered.
While most approaches bear the necessity of an oscillating magnetic field which can be tuned into the frequency of the time-fluctuating spin moment \cite{manassen1989,manassen1997,durkan2002,xiao2004,koppens2006,epstein2005}, ESR has been achieved in absence of such field \cite{pioro2008,Science.350.417}. The use of oscillating magnetic fields is a great disadvantage since generating strong and localized magnetic fields, necessary for addressing single spins, is technically challenging. Ways to circumvent the difficulties associated with high frequency electromagnetic field generation were exploited in \cite{billangeon2007,bretheau2013}, where high frequency photons were generated in Josephson junction design.

Motivated by the recent experimental progress reported in \cite{Science.350.417}, in this article we propose a different set-up in which only a static magnetic field is necessary whereas an external frequency is brought into the system through a time-dependent electric field.
This technique is available for spin polarized tunneling currents which generate an asymmetry in the spin resolved conductance channels that is sensitive to low energy fluctuations in the localized magnetic moment that is embedded in the tunnel junction. As the tunneling electrons couple to the localized magnetic moment via exchange, the frequencies of the temporal spin fluctuations in the molecule mix with the frequency of the electric field, and through this coupling the spin polarized current picks up the frequencies corresponding to the low energy spectrum of the magnetic sample. We expect that this approach is applicable both in conventional break junctions with spin polarized leads and spin polarized (SP) STM.

The electric field that is employed as a driving source for the spin transitions does not provide spin angular momentum to the system. Therefore, an ESR signal can only be measured for spins in which transitions between the ground and first excited states that are spin angular momentum conservative. We show that the spin polarized current itself generates the corresponding transverse anisotropy field which is sufficient to support such transitions and, hence, an ESR signal for integer (1, 3, $\ldots$) spins. We also explain why this field is not sufficient to generate an ESR signal for half integer (1/2, 3/2, $\ldots$) spins. Our proposed model is, therefore, capable of explaining both the ESR measurements using SP-STM \emph{and} the different results on Fe ($S=2$) and Co ($S=3/2$) observed in Ref. \cite{Science.350.417}.

It is important to point out that the effect predicted in this article generates a different type of ESR compared to conventional approaches. Typically, ESR is considered as noise spectroscopy for transitions between different spin states, such that the static field induced Zeeman split is detuned by the frequency of an oscillating field that provides a coupling between the spin states. Here, we show that the spin polarized tunneling current comprise a component proportional to  $\av{S_z}$ such that ESR between the ground and first excited states are picked up directly as a time-dependent component in the total current. The effect may, therefore, be used to probe the low energy spin states in molecular magnetic compounds, e.g., Cr$_8$, Cr$_7$Ni, Fe$_4$, \cite{chiesa2013} or paramagnetic $M$-phthalocyanine, where M denotes a transition metal element \cite{coronado2004,chen2008,mugarza2012,krull2013}, and other suitable compounds \cite{Science.350.417,coronado2004}.
%

\begin{figure}[t]
\begin{center}
\includegraphics[width=0.99\columnwidth]{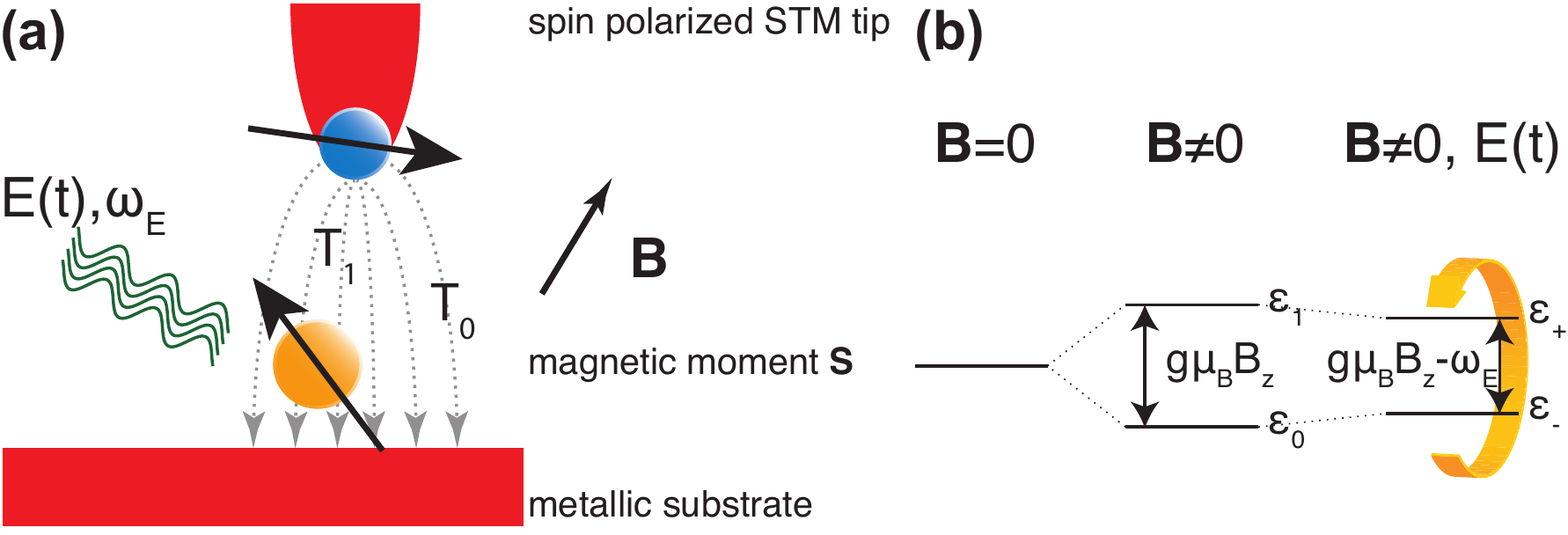}
\end{center}
\caption{
{\bf Set-up for ESR using SP-STM and time-dependent electric field.}
{\bf (a)} A spin polarized STM set-up with a magnetic tip and localized paramagnetic moment $\bfS$. A static magnetic field $\bfB$ generates a spin polarization in the local moment. The electric field ${\bf E}$ drives the spin resonance with detuning frequency $\omega_E$.
{\bf (b)} Schematics of a two-level system under vanishing magnetic field ($\bfB=0$) for which the levels are degenerate, and under finite magnetic field ($\bfB\neq0$) for which the levels are split by $g\mu_BB_z$. Application of the electric field ${\bf E}(t)$ generates a coupling between the states and the effective level separation is controlled via the detuning frequency $\omega_E$.}
\label{fig-Fig1}
\end{figure}

\section*{RESULTS}
\subsection*{Time-dependent tunneling current}
For the sake of argument and in order to demonstrate the gross effect, we derive the tunneling current within a model comprising the salient features of the physics we consider. The set-up consists of a single spin $\bfS(t)$ embedded in the tunnel junction between two metallic leads, see Fig. \ref{fig-Fig1} (a) for schematics. We model the system through the Hamiltonian
\begin{align}
\Hamil=&
	\Hamil_0+\Hamil_S+\Hamil_T,
\label{eq-Hamiltonian}
\end{align}
where $\Hamil_0=\sum_{\bfk\sigma\in L,R}\leade{\bfk}\cdagger{\bfk}\cc{\bfk}$ denotes the Hamiltonian for the left ($L$) and right ($R$) lead, whereas $\cdagger{\bfk}$ ($\cc{\bfk}$) creates (annihiliates) an electron at the energy $\leade{\bfk}$ with momentum $\bfk$ and spin $\sigma=\up,\down$. The contribution $\Hamil_S$ defines the model for the local spin and will be discussed in more detail later. Tunneling between the leads in presence of the local spin is modeled by $\Hamil_T=\sum_{\bfp\bfq\sigma\sigma'}\cdagger{\bfp}\hat{\bfT}_{\sigma\sigma'}\cs{\bfq\sigma'}+H.c.$, where we use $\bfp$ ($\bfq$) for states in the left (right) lead and where $\hat\bfT_{\sigma\sigma'}=T_0\sigma^0_{\sigma\sigma'}+T_1\bfS\cdot\bfsigma_{\sigma\sigma'}$. Here, $\bfsigma$ and $\sigma^0$ are the Pauli matrix vector and identity matrix, respectively.

We derive the tunneling current up to quadratic order in the tunneling rate. Hence, the time-dependent tunneling current $I(t,V)$ across the junction can be written
\begin{align}
I(t,V)=&
	-\frac{2e}{\hbar}\im
	(-i)\int_{-\infty}^t
			\av{\com{A(t)}{A^\dagger(t')}}e^{-ieV(t-t')}
	dt',
\label{eq-Idef}
\end{align}
where the operator $A(t)=\sum_{\bfp\bfq\sigma\sigma'}\cdagger{\bfp}(t)\hat\bfT_{\sigma\sigma'}(t)\cs{\bfq\sigma'}(t)$.
The functional form of the tunneling matrix $\hat\bfT$ allows for partition of the current into three components $I=\sum_{n=1,2,3}I^{(n)}$ \cite{NanoLett.9.2414,PhysRevB.81.115454}, each of which represents a different tunneling processes. The first component, $I^{(1)}\propto T_0^2$, does not couple directly to $\bfS$, but merely provides a stationary back-ground current. The third contribution, $I^{(3)}\propto T_1^2$, provides a coupling to the spin noise $\av{\bfS(t)\bfS(t')}$, which has been extensively discussed previously 
\cite{manassen1989,manassen1997,messina2007,mannini2007,komeda2008,manassen2014,QIP.1.355,PhilMagB.82.1291,PhysRevB.66.195416,durkan2002,durkan2004,manniniLang2007,krukowski2010,balatsky2012}. 
Under stationary voltage bias, however, this component is also stationary. Hence, as these contributions are stationary, they will be omitted in the following discussion.

In contrast, the second contribution, $I^{(2)}(t,V)\propto T_0T_1$, contains a direct coupling to the local spin and its dynamics. We write this contribution as
\begin{align}
I^{(2)}_J(t,V)=&
	\frac{2e}{h}
	\im\,
	\int
		\langle
			S_z(t)
			+S_z(t')
		\rangle
		\Phi(t,t')
	dt'
	,
\label{eq-I2t}
\end{align}
where $\Phi(t,t')$ describes the correlations between electrons tunneling through the junction \cite{PhysRevB.81.115454}.

Notice that our formulation of the ESR differ from previous studies \cite{QIP.1.355,PhilMagB.82.1291,PhysRevB.66.195416,balatsky2012} at this point, since we go beyond the adiabatic approximation for the spin even though the time-scales of the spin and electronic degrees of freedom may be significantly different. Hence, by taking into account the full time-evolution of the local spin, it becomes obvious from Eq. (\ref{eq-I2t}) that the time-evolution of the tunneling current directly depends on the dynamics of the local spin moment. As we shall see below, this current component is modulated by the precession of the local spin.

We can obtain a simple estimate of the expected time-dependent contribution to the current by neglecting the back-action from the localized spin on the tunneling electrons. Hence, the electronic degrees of freedom become time-independent and we can integrate out the time variable $t'$ and write the current $I^{(2)}$ as
\begin{align}
I^{(2)}(t,V)=&
	\frac{2e}{h}
	\im
		\int
			\av{S_z(\omega)}
			{\cal F}(\omega,V)
			e^{-i\omega t}
		d\omega
	,
\label{eq-I2}
\end{align}
where ${\cal F}(\omega,V)=\Phi(eV)+\Phi(eV-\omega)$, with
\begin{align}
\Phi(\dote{})=&
	-T_0T_1
	\sum_{\bfp\bfq\sigma}\sigma^z_{\sigma\sigma}
			\frac{f(\leade{\bfp})-f(\leade{\bfq})}{\leade{\bfp}-\leade{\bfq}-\dote{}+i\delta}
		,
\label{eq-Phi}
\end{align}
where $f(x)$ is the Fermi function, whereas $\delta>0$ is infinitesimal.

The current given in Eq. (\ref{eq-I2}) includes a time-dependence which involve the temporal fluctuations of the local spin moment. This is significant since it provides a convolution between the local spin moment and the density of tunneling electron states, which open for the opportunity to tune in the voltage bias near the spin excitations and into a regime with a resonant tunneling current. The presence of the Pauli matrix $\sigma^z_{\sigma\sigma}$ in Eq. (\ref{eq-Phi}) emphasizes that this time-dependent contribution to the tunneling current is non-vanishing only whenever there is an asymmetry between the spin-channels in the system, i.e. a finite spin-polarization. This is most easily seen by converting the momentum summations to energy integrations over the spin-resolved densities of electron states $n_\sigma$ and $N_\sigma$ in the tip and substrate, respectively, which are assumed to have a slow energy variation. By defining $n_\sigma=n_0(1+\sigma^z_{\sigma\sigma}p_t\cos\theta)/2$ ($N_\sigma=N_0(1+\sigma^z_{\sigma\sigma}P_s)/2$), we obtain $\Phi(\dote{})\sim\sum_\sigma\sigma^z_{\sigma\sigma}n_\sigma N_\sigma=n_0N_0(p_t\cos\theta+P_s)/2$. Here, the total density of electron states and the spin polarization in the tip (substrate) are denoted by $n_0$ ($N_0$) and $p_t$ ($P_s$) $\in(0,1)$, respectively, whereas $\theta$ defines the angle between the spin quantization axes of the tip and substrate. The current is proportional to the sum of the spin polarizations in the tip and substrate. Therefore, a necessary condition for the current in Eq. (\ref{eq-I2}) to be finite is that at least one electrode supports spin polarized electrons and that $P_s\neq-p_t\cos\theta$. 

\begin{figure}[t]
\begin{center}
\includegraphics[width=0.99\columnwidth]{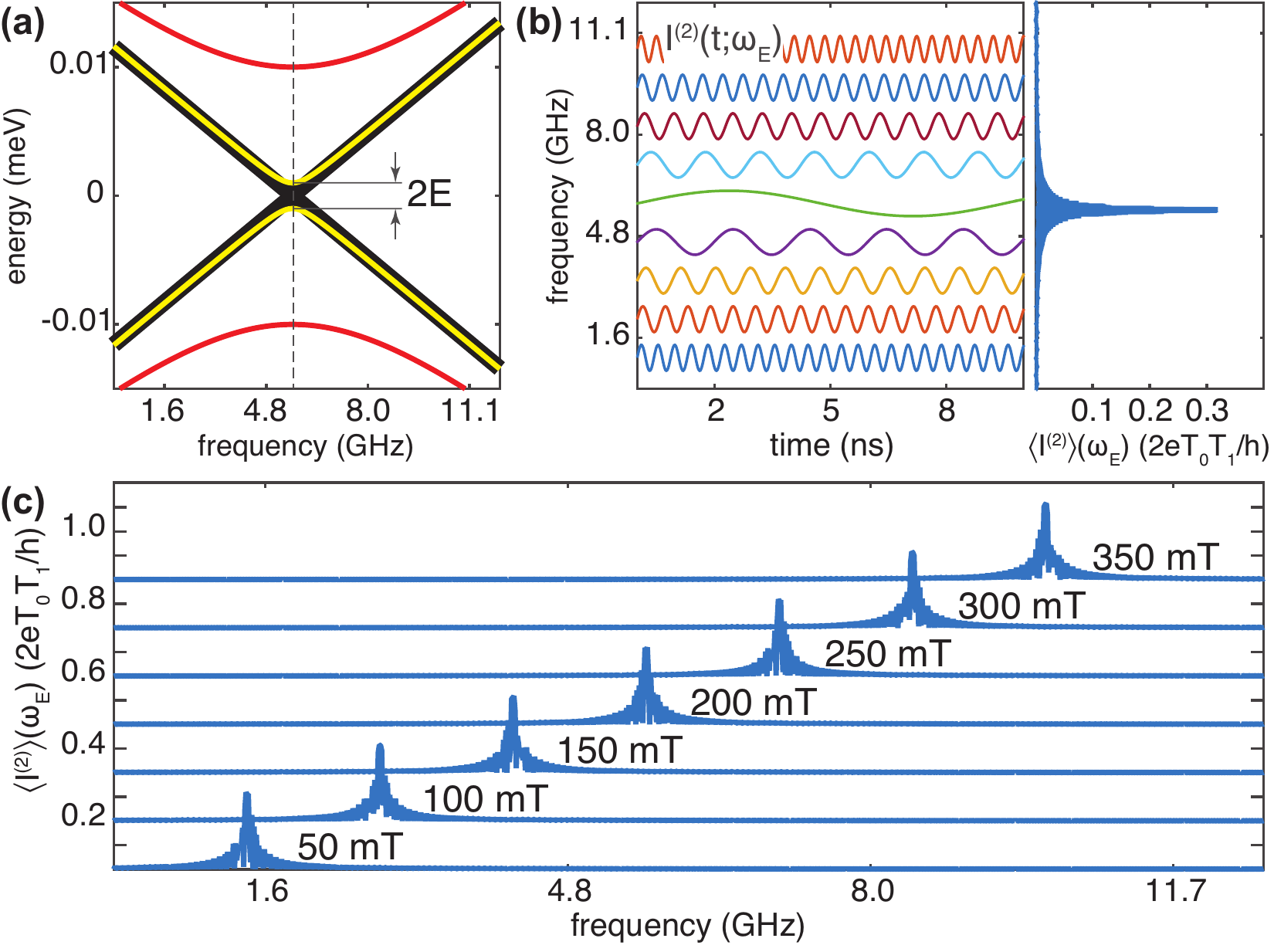}
\end{center}
\caption{
{\bf Frequency dependence of energy spectrum and transport data.}
(a) Eigenenergies $\dote{\pm}$ as function of $\omega_E$ for different electric field strengths $E=0.1$ $\mu$eV (black), $1$ $\mu$eV (yellow), and $10$ $\mu$eV (red) at $B=200$ mT.
(b) Corresponding current $I^{(2)}(t)$ as function of time (left) for different frequencies $\omega_E$ and time averaged current $\av{I^{(2)}}(\omega_E)$ as function of $\omega_E$ in the case $E=0.1$ $\mu$eV.
(c) Time averaged current $\av{I^{(2)}}(\omega_E)$ using $E=0.1$ $\mu$eV in increasing order for the magnetic field strengths $B_z\in(50,350)$ mT with increments of $50$ mT. Traces are off-set for clarity.
Other parameters are $n_0=N_0=1$, $p_t=1/2$, $P_s=0$, and $T=0.6$ K.}
\label{fig-Fig2}
\end{figure}

\subsection*{Two-level system}
Next, we consider a simplified example of the ESR using an external electric field applied to a localized spin moment, defined by a degenerate two-level system with the states and energies $\{\ket{n},\ \dote{n}\}$, $n=1,2$, where $\dote{n}=\dote{0}$. We can write the spin Hamiltonian $\Hamil_S=\sum_n[\dote{0}-(-1)^n\omega_r/2]n_n+E(\ddagger{2}\dc{1}e^{-i\omega_Et}+H.c.)$, where $\omega_r=g\mu_BB_z$ defines the resonance energy.  Here, $\ddagger{n}$ ($\dc{n}$) creates (annihilates) a particle in the state $\ket{n}$ and $n_n=\ddagger{n}\dc{n}$, whereas $B_z$ is a static external magnetic field, $g$ is the gyromagnetic ratio, $\mu_B$ is the Bohr magneton, and $E$ is the effective coupling between the states $\ket{1}$ and $\ket{2}$ provided by the electric field with frequency $\omega_E$. Without loss of generality we can assume that $\dote{0}=0$. The system is transformed into the rotating reference frame of the electric field through $\tilde\Hamil_S=e^{\calS}\Hamil_Se^{-\calS}+i(\dt e^{\calS})e^{-\calS}$, with $\calS=-i(\omega_Et/2)(n_1-n_2)$, in order to eliminate the time-dependence from the Hamiltonian at the cost of introducing the energy shift $(-1)^n\omega_E/2$ to the energy $\dote{n}$. The eigenstates of the resulting model are given by $\ket{\pm}=\alpha_\pm\ket{0}+\beta_\pm\ket{1}$, with corresponding eigenenergies $\dote{\pm}=\pm\sqrt{(\omega_r-\omega_E)^2+4E^2}/2$.
Fig. \ref{fig-Fig1} (b) illustrates how the spin states of the local moment split up under application of an external static magnetic field and an external fluctuating electric field while
the plots in Fig. \ref{fig-Fig2} (a) show $\dote{\pm}$ as function of $\omega_E$ for different electric field strengths $E=0.1$ $\mu$eV (black), $1$ $\mu$eV (yellow), and $10$ $\mu$eV (red) at $B_z=200$ mT.

In the eigenstate representation we can write $\av{S_z(\omega)}=\sum_{s=\pm}\sigma^z_{ss}\calP(\omega)\delta(\omega-\dote{s})$, where $\calP(\omega)$ defines the distribution of the density of occupied states in the two-level system. Inserting the expression for $\av{S_z(\omega)}$ into the current, Eq. (\ref{eq-I2}), yields
\begin{align}
I^{(2)}(t,V)=&
	\frac{2e}{h}
	\im
	\sum_{s=\pm}\sigma_{ss}^z\calP(\dote{s}){\cal F}(\dote{s},V)e^{-i\dote{s}t}
	.
\label{eq-I2tls}
\end{align}
This contribution provides a time-dependent current with the characteristic frequencies $\dote{\pm}$ which are mixtures of the intrinsic energies of the two-level system and the parameters of the external electric field. The finiteness of this current crucially relies on the inequality $\calP(\dote{+})\neq\calP(\dote{-})$ which is, typically, fulfilled whenever the states are non-degenerate.
By tuning the frequency of the electric field into resonance $\omega_E\rightarrow\omega_r$, such that $\sqrt{(\omega_r-\omega_E)^2+4E^2}\approx2E$, see Fig. \ref{fig-Fig2} (a), the frequency of the current $I^{(2)}(t,V)$ is minimized and for small coupling $E$, the current becomes nearly constant, see Fig. \ref{fig-Fig2} (b).

Experimental resolution of the high frequency oscillations in the tunneling current generally presents a great challenge and it is often more convenient to measure the time-averaged current $\av{I^{(2)}}(V)=\lim_{\calT\rightarrow\infty}\int_{-\calT/2}^{\calT/2}I^{(2)}(t,V)dt/\calT$. Within the two-level system we obtain
\begin{align}
\av{I^{(2)}}(V)=&
	\frac{2e}{h}
	\lim_{\calT\rightarrow\infty}
	\im
		\sum_{s=\pm}
			\sigma^z_{ss}
			\calP(\dote{s})
			{\cal F}(\dote{s},V)
			\frac{\sin\dote{s}\calT/2}{\dote{s}\calT/2}
	.
\label{eq-I2av}
\end{align}
This average is finite only for $\dote{s}\approx0$, which corresponds to the presence of a constant term in Eq. (\ref{eq-I2tls}), see Fig. \ref{fig-Fig2} (b) where we plot $\av{I^{(2)}}(V;\omega_E)$. Hence, by tuning the frequency $\omega_E$ into resonance the total current is increased roughly by $2eT_0T_1\sum_s\sigma^z_{ss}\calP(E){\cal F}(E)/h$, see Fig. \ref{fig-Fig2} (b). The traces in Fig. \ref{fig-Fig2} (c) show $\av{I^{(2)}}(V;\omega_E)$ for different magnetic fields $B_z$ (see figure caption for details) and the plots clearly demonstrate the linear shift of the resonance frequency, in excellent agreement with the results in \cite{Science.350.417}.

\section*{DISCUSSION}
In many studies of localized spin interacting with tunneling currents, the intrinsic spin Hamiltonian is assumed to be on the form \cite{NanoLett.9.2414,PhysRevB.81.115454,PhysRevLett.102.256802,PhysRevLett.103.050801,PhysRevLett.103.176601}
\begin{align}
\Hamil_S=&
	-g\mu_B\bfB\cdot\bfS+DS_z^2+E(S_+^2+S_-^2)/2
	,
\label{eq-HS}
\end{align}
where $D$ and $E$ represent the uniaxial and transverse anisotropy fields, respectively, whereas $\bfB$ is the effective magnetic field. Here, we show that this model can be justified as a result of interactions between the localized spin and the electrons in the substrate (and in the tip) as well as from the interactions between the localized spin and the tunneling current. We also show that these contributions to the anisotropies can be controlled by the voltage bias and the distance between the tip and the sample, where the latter effect may be viewed in perspective of the results in \cite{NatPhys.9.765,NanoLett.15.4024}

Starting from the model given in Eq. (\ref{eq-Hamiltonian}), we construct an effective model for the local spin on the Keldysh contour in order to account for the non-equilibrium conditions in the system. By integrating out the Fermionic degrees of freedom \cite{PhysRevLett.92.107001,NewJPhys.10.013017,PhysRevLett.108.057204,PhysRevLett.113.257201}, the pertinent effective spin action for this interaction assumes the form
\begin{align}
\calS_\text{eff}=&
	-\frac{1}{2}
	\oint
		\hat{\bfT}_{\sigma\sigma'}(t)
			\calD_{\sigma\sigma'}(t,t')
		\hat{\bfT}_{\sigma'\sigma}(t')
	dtdt',
\label{eq-S}
\end{align}
where the current-current propagator $\calD_{\sigma\sigma'}(t,t')=(-i)\av{{\rm T}A_{\sigma\sigma'}(t)A^\dagger_{\sigma'\sigma}(t')e^{ieV(t-t')}+A_{\sigma\sigma'}^\dagger(t)A_{\sigma'\sigma}(t')e^{-ieV(t-t')}}$, and $A_{\sigma\sigma'}(t)=\sum_{\bfp\bfq}\cdagger{\bfp}(t)\cs{\bfq\sigma'}(t)$.

By grouping into three contributions, one finds a term proportional to $T_0^2$ which does not couple to the spin and will, therefore, be omitted. The other two terms, which are proportional to $T_0T_1$ and $T_1^2$, respectively, provide (i) a current induced magnetic field ($B_t$), and (ii) current induced uniaxial ($D_t$) and transverse ($E_{ij}$, $i,j=x,y$) anisotropy fields acting on the local spin. We analyze the effect of the two anisotropy fields by mapping this model onto an effective Hamiltonian on the form [In principle the is also a contribution of the type $\calE\cdot(\bfS\times\bfS)$ which, however, vanishes identically for a single spin.]
\begin{align}
\Hamil_t=&
	-g\mu_BB_tS_z
	+D_t S_z^2
	+\sum_{ij=xy}S_iE_{ij}S_j
	,
\label{eq-Ht}
\end{align}
where the magnetic, or, fine structure field $B_t=-T_0T_1\sum_\sigma\sigma^z_{\sigma\sigma}\calN_{\sigma\sigma}(V)/g\mu_B$ with $\calN_{\sigma\sigma'}(V)=\oint\calD_{\sigma\sigma'}(t,t')dt'$, whereas the uniaxial anisotropy $D_t=-T_1^2\sum_\sigma\calN_{\sigma\sigma}(V)/2$ and transverse anisotropy $E_{ij}=-T_1^2\sum_\sigma\sigma^i_{\sigma\bar\sigma}\calN_{\sigma\bar\sigma}(V)\sigma^j_{\bar\sigma\sigma}/2$.

Before we discuss the properties of these fields, we show how the model in Eq. (\ref{eq-Ht}) can be turned into the form represented by $\Hamil_S$ in Eq. (\ref{eq-HS}). Noticing that $E_{xy}=-E_{yx}\equiv i\calE_\perp$ and $E_{xx}=E_{yy}\equiv\calE_{\|}$, we rotate the $xy$-plane using the unitary transformation $(\tilde{S}_x \ \tilde{S}_y )=(S_x \ S_y)(\sigma_0-i\sigma_x)/\sqrt{2}$ which enables us to write the model $\Hamil_t$ as
\begin{align}
\Hamil_t=&
	-
		g\mu_BB_tS_z
	+
		(D_t-\calE_{\|})S_z^2
	+
		\calE_\perp(\tilde{S}_x^2-\tilde{S}_y^2)
	+
		\calE_{\|}S^2.
\label{eq-effHt}
\end{align}
Here, the last term merely provides a constant shift of the excitation spectrum and is therefore discarded. Finally, by switching to the ladder operators $S_\pm=\tilde{S}_x\pm i\tilde{S}_y$, we retain the form given in Eq. (\ref{eq-HS}).

We proceed by considering the properties of the current induced fields $B_t$, $D_t$, and $E_{ij}$. For simplicity, we approximate the current propagator by decoupling into single electron Green functions (GFs) according to $\calD_{\sigma\sigma'}(t,t')=(-i)\sum_{\bfp\bfq}G_{\bfq\sigma'}(t,t')G_{\bfp\sigma}(t',t)$, where $G_{\bfk\sigma}(t,t')=\eqgr{\cc{\bfk}(t)}{\cdagger{\bfk}(t')}$ is the GF for the lead $\chi=L,R$. We also assume that back-action effects from the localized spin moment on the tunneling electrons are negligible.

For stationary bias voltages the integrals in the current induced fields $B_t$, $D_t$, and $E_{ij}$ can be evaluated. In particular, for collinear spin polarized leads we have
\begin{align}
\calN_{\sigma\sigma'}=&
	\sum_{\bfp\bfq}
	\biggl(
		\frac{f(\leade{\bfp})-f(\dote{\bfq\sigma'})}{\leade{\bfp}-\dote{\bfq\sigma'}-eV+i\delta}
		+
		\frac{f(\dote{\bfp\sigma'})-f(\leade{\bfp})}{\dote{\bfp\sigma'}-\leade{\bfp}+eV-i\delta}
	\biggr).
\end{align}
In order to estimate the effect of the induced fields, we treat the summations and densities of electron states in the tip and substrate as above. We obtain
\begin{align}
\calN_{\sigma\sigma'}\approx&
	\frac{n_0N_0}{2}\calQ(V)
	(1+\sigma^z_{\sigma\sigma}p_t\cos\theta)(1+\sigma^z_{\sigma'\sigma'}P_s)
	.
\end{align}
Here, the real part of $\calQ(V)\sim\int_{|\dote{}|<D_t}f(\dote{})\{\ln|[(\dote{}-D_s)^2-(eV)^2]/[(\dote{}+D_s)^2-(eV)^2]|+i\pi[\theta(\dote{}-eV)-\theta(\dote{}+eV)]\}d\dote{}$ essentially depends on the band width $D_t$ ($D_s$) of the metallic tip (substrate) while the imaginary part depends linearly on the voltage bias $V$ across the junction. We notice that the induced fields depend on the densities of electron states in the tip ($n_0$) and substrate ($N_0$) as well as the voltage bias across the junction. More important, however, is the strong dependence on their respective spin polarization. For instance, the induced magnetic field $B_t(V)\propto(p_t\cos\theta+P_s)\calQ(V)$ is finite for spin polarized currents, unless the tip and substrate are equally spin polarized but in anti-parallel orientation ($P_s=-p_t\cos\theta$). While the uniaxial and transverse anisotropy fields $D_t(V)\propto(1+p_tP_s\cos\theta)\calQ(V)$ and $\calE_{\|}\propto(1-p_tP_s\cos\theta)\calQ(V)$, respectively, are finite for all non-equilibrium conditions [This is true for $D_t$ ($\calE_{\|}$) except in the extreme case with half-metallic tip and substrate in anti-parallel (parallel) configuration, e.g., $P_s=-p_t\cos\theta=1$ ($P_s=p_t\cos\theta=1$).], the total uniaxial anisotropy $(D_t-\calE_{\|})(V)\propto p_tP_s\calQ(V)\cos\theta$, c.f. Eq. (\ref{eq-effHt}), is finite only when both the tip and substrate are spin polarized.
Finally, the transverse anisotropy field $\calE_\perp(V)\propto(p_t\cos\theta-P_s)\calQ(V)$ is finite for spin polarized currents except when the tip and substrate are spin polarized equally and in parallel configuration ($P_s= p_t\cos\theta$).

The above discussion can be equally applied to the (exchange) interactions between the localized spin and the electrons in the tip (substrate). By generalizing the derivation in, e.g., Refs. \onlinecite{PhysRevLett.108.057204,PhysRevLett.113.257201,PhysRevB.92.125405} one finds that the resulting anisotropy fields (i) are finite only in materials with non-trivial magnetic structure, e.g., finite spin polarization and/or spin chirality (spin-orbit coupling) and (ii) can be summarized in a model on the form given in Eq. (\ref{eq-HS}). However, for finite spin-orbit coupling terms proportional to $S_xS_z$, $S_yS_z$, et c, contribute to the model.

Regarding the influence of the electric field, we notice that an electric field ${\bf E}(t)$ acts on the local spin according to $[\hat{\bf z}\times{\bf E}(t)]\cdot\bfS$. Using the procedure that was employed to derive Eqs. (\ref{eq-HS}) and (\ref{eq-S}) on this contribution, we obtain an effective Hamiltonian on the form $\Hamil_1(t)=V_+(t)S_+^2+V_-(t)S_-^2+V_\perp(t)(S_+S_-+S_-S_+)$, where $V_{+/-/\perp}(t)$ describe different combinations of the $x$- and $y$-components of ${\bf E}(t)$.

In the experiments reported in Ref. \onlinecite{Science.350.417}, a spin polarized tip is used to measure the response of adsorbed Fe and Co atoms, and while the MgO substrate lacks magnetization it may provide a finite spin-orbit coupling. The experimental results show that the ESR which was observed for Fe is completely absent for Co. We propose an explanation for this different behavior based on the anisotropy fields induced by the polarized tunneling current.
As the spin polarization $P_s$ in the substrate is negligible, the above discussion suggests that the uniaxial anisotropy $D-\calE_{\|}$ induced from the tunneling current vanishes, while the transverse field $\calE_\perp$ is finite. Employing Eq. (\ref{eq-effHt}) to a spin $S=2$, pertaining to Fe adsorbed onto MgO \cite{Science.350.417,PhysRevLett.115.237202}, assuming a negative uniaxial anisotropy $D$ induced by the coupling to the substrate, we find that the ground and first excited states are given as the superpositions
\begin{align}
\ket{n}=&
	\alpha_n\ket{2,2}
	+
	\beta_n\ket{2,0}
	+
	\gamma_n\ket{2,-2}
	,
	\
	n=0,1,
\end{align}
where the coefficients $\alpha_n$, $\beta_n$, and $\gamma_n$ depend on the parameters of the model. Here, the eigenstates are expressed in terms of the Fock basis $\ket{S,m_z}$. The transition matrix element between the ground and first excited states induced by the electric field, $\bra{1}\Hamil_1(t)\ket{0}$, is in this case finite. The transitions are, hence, accessible through the ESR measurement, despite no spin angular momentum in the $z$-direction of the Fock basis is provided. In the case of a spin $S=3/2$, which is relevant for Co adsorbed onto MgO \cite{Science.344.988}, the situation is quite different. As ground and first excited states we obtain
\begin{align}
\ket{0}=&
	\alpha_0\ket{3/2,3/2}
	+
	\beta_0\ket{3/2,-1/2}
	,
\\
\ket{1}=&
	\alpha_1\ket{3/2,-3/2}
	+
	\beta_1\ket{3/2,1/2}
	.
\end{align}
Notice that these states do not share the same Fock states which implies that for spin transitions to take place, spin angular momentum in the $z$-direction of the Fock basis has to be provided by either the external time-dependent field or by the tunneling current. However, the current contribution we discuss in the present paper does not support any exchange of spin angular momentum between the current and the localized spin moment, hence, it can only be provided by the external source. A linearly polarized electric field does not provide the necessary spin angular momentum which means that no ESR can be achieved, which is also verified by the vanishingly small transition matrix element $\bra{1}\Hamil_1(t)\ket{0}$ in this case. We therefore conjecture that the current induced anisotropies are sufficient to generate the electric field controlled ESR for localized moments with integer spins (1, 2, $\ldots$) but not with half integer spins (1/2, 3/2, $\ldots$).

The existence of the anisotropy fields exerted by the tip (substrate) and tunneling current on the local spin moment in presence of spin polarization, opens for controlled manipulations of the spectral details of the localized spin moment, in analogy to the measurements on local spin moments using superconducting STM \cite{NatPhys.9.765,NanoLett.15.4024}. As the exchange interaction between the spin and the electrons in the tip (substrate) depends exponentially on the distance between the tip (substrate) and the sample, based on our previous results \cite{EPL.108.67009,PhysRevB.91.205438} we predict that the ESR frequency shifts as a function of the distance between the tip (substrate) and the sample. Experimentally, this is likely to be verified most easily by varying the distance of a spin polarized tip relative to the sample. The resulting increased anisotropy then generates a redistributed spin excitation spectrum which accordingly changes the resonance frequency.

\section*{CONCLUSIONS}
In conclusion, we have introduced a theoretical tool for ESR using spin polarized STM and an external time-dependent electromagnetic field. We show that ESR in this configuration is possible only under spin polarized conditions since the spin asymmetry is required in order to probe local spin fluctuations. Furthermore, we show that the spin polarized conditions in the system are sufficient to generate finite uniaxial and transverse anisotropy fields as well as a current induced magnetic field, which act on the local spin moment. These fields are sufficient to support electric field induced ESR between the ground and first excited states for integer spin moments, while no ESR signal is expected in this set-up for half integer spins. Our results are in excellent agreement with the experimental observations of ESR reported in Ref. \onlinecite{Science.350.417}. We finally predict that the strengths of the anisotropy fields depend on the distance between the tip (substrate) and the sample, which opens for controlled manipulations of the spin excitation spectrum. Accordingly, the ESR frequency is expected to shift as function of this distance.

\section*{Acknowledgements}
We gratefully thank A. V. Balatsky, A. Bergman, K. J. Franke, L. Nordstr\"om, J. Nilsson, J. I. Pascual, H. Ottosson, and M. Ternes, for stimulating and fruitful discussions. Support from the Swedish Research Council is acknowledged.


\begin{thebibliography}{20}

\bibitem{manassen1989} Manassen, Y., Hamers, R. J., Demuth, J. E., \& Castellano Jr., A. J., Direct observation of the precession of individual paramagnetic spin on oxidized silicon surfaces, \emph{Phys. Rev. Lett.} {\bf 62}, 2531--2534 (1989).
\bibitem{manassen1997} Manassen, Y., Real-Time Response and Phase-Sensitive Detection to Demonstrate the Validity of ESR-STM Results, \emph{J. Magn. Reson.} {\bf 126}, 133--137 (1997).
\bibitem{messina2007} Messina, P. et al.
Spin noise fluctuations from paramagnetic molecular adsorbates on surfaces, \emph{J. Appl. Phys.} {\bf 101}, 053916 (2007).
\bibitem{mannini2007} Mannini, M. et al.
Addressing individual paramagnetic molecules through ESN-STM, \emph{Iorganica Chimica Acta}, {\bf 360}, 3837--3842 (2007).
\bibitem{komeda2008} Komeda, T. \& Manassen, Y., Distribution of frequencies of a single precessing spin detected by scanning tunneling microscope, \emph{Appl. Phys. Lett.} {\bf 92}, 212506 (2008).
\bibitem{manassen2014} Manassen, Y., Averbukh, M., \& Morgenstern, M., Analyzing multiple encounter as a possible origin of electron spin resonance signals in scanning tunneling microscopy on Si(111) featuring C and O defects, \emph{Surf. Sci.} {\bf 623}, 47 (2014).

\bibitem{QIP.1.355} Balatsky, A.V. \& Martin, I., Theory of Single Spin Detection with STM, \emph{Quant. Inform. Process.} {\bf 1}, 355--364  (2002).
\bibitem{PhilMagB.82.1291} Balatsky, A.V., Manassen, Y., \& Salem, R., Exchange-based noise spectroscopy of a single precessing spin with scanning tunnelling microscopy, \emph{Phil. Mag. B}, {\bf 82}, 1291--1298 (2002).
\bibitem{PhysRevB.66.195416} Balatsky, A.V., Manassen, Y., \& Salem, R., ESR-STM of a single precessing spin: Detection of exchange-based spin noise, \emph{Phys. Rev. B}, {\bf 66}, 195416 (2002).

\bibitem{durkan2002} Durkan, C. \& Welland, M.E., Electronic spin detection in molecules using scanning-tunneling- microscopy-assisted electron-spin resonance, \emph{Appl. Phys. Lett.} {\bf 80}, 458--460 (2002).
\bibitem{durkan2004} Durkan, C., Detection of single electronic spins by scanning tunnelling microscopy, \emph{Contemp. Phys.} {\bf 45}, 1--10 (2004).
\bibitem{manniniLang2007} Mannini, M. et al.
Self-Assembled Organic Radicals on Au(111) Surfaces: A Combined ToF-SIMS, STM, and ESR Study, \emph{Langmiur}, {\bf 23}, 2389--2397 (2007).
\bibitem{krukowski2010} Krukowski, P. et al.
An ESN-STM spectrometer for single spin detection, \emph{Measurement}, {\bf 43}, 1495--1502 (2010).
\bibitem{biointerfaces.10.031001} Naruszewicz, M. et al.
Detection and analysis of spin signal in spin-labeled poly(l-lysine), \emph{Biointerfaces}, {\bf 10}, 031001 (2015).

\bibitem{balatsky2012} Balatsky, A.V., Nishijima, M., \& Manassen, Y., Electron spin resonance-scanning tunneling microscopy, \emph{Adv. Phys.} {\bf 61}, 117--152 (2012).

\bibitem{xiao2004} Xiao, M., Martin, I., Yablonovitch E., \& Jiang H. W., Electrical detection of the spin resonance of a single electron in a silicon field-effect transistor, \emph{Nature}, {\bf 430}, 435--439 (2004).
\bibitem{koppens2006} Koppens, F.H.L. et al.
Driven coherent oscillations of a single electron spin in a quantum dot, \emph{Nature}, {\bf 442}, 766--771 (2006).
\bibitem{pioro2008} Pioro-Ladri\'ere, M. et al.
Electrically driven single-electron spin resonance in a slanting Zeeman field, \emph{Nature Phys.} {\bf 4}, 776--779 (2008).

\bibitem{hao2014} Hao, X., Ruskov, R., Xiao, M., Tahan, C., \& Jiang, H.-W., Electron spin resonance and spin-valley physics in a silicon double quantum dot, \emph{Nature Comm.} {\bf 5}, 3860 (2014).

\bibitem{epstein2005} Epstein, R.J., Mendoza, F.M., Kato, Y.K., \& Awschalom, D.D., Anisotropic interactions of a single spin and dark-spin spectroscopy in diamond, \emph{Nature Phys.} {\bf 1}, 94--98 (2005).

\bibitem{molotkov1992} Molotkov, S.N., On the theory of the tunneling current modulation at the Larmor frequency due to precession of an individual spin on a paramagnetic center, \emph{Surf. Sci.} {\bf 264}, 235 (1992).
\bibitem{prioli1995} Prioli, R. \& Helman, J.S., Effect of resonating paramagnetic centers on the current of the scanning tunneling microscope, \emph{Phys. Rev. B} {\bf 52}, 7887 (1995).
\bibitem{engel2001} Engel, H.-A. \& Loss, D., Detection of Single Spin Decoherence in a Quantum Dot via Charge Currents, \emph{Phys. Rev. Lett.} {\bf 86}, 4648--4651 (2001).
\bibitem{mozyrsky2002} Mozyrsky, D., Fedichkin, L., Gurvitz, S.A., \& Berman, G.P., Interference effects in resonant magnetotransport, \emph{Phys. Rev. B} {\bf 66}, 161313(R) (2002).
\bibitem{zhu2002} Zhu, J.-X. \& Balatsky, A.V., Quantum Electronic Transport through a Precessing Spin, \emph{Phys. Rev. Lett.} {\bf 89}, 286802 (2002).
\bibitem{levitov2003} Levitov, L.S. \& Rashba, E.I., Dynamical spin-electric coupling in a quantum dot, \emph{Phys. Rev. B}, {\bf 67}, 115324 (2003).
\bibitem{bulaevskii2003} Bulaevskii, L.N., Hruska, M., \& Ortiz, G., Tunneling measurement of quantum spin oscillations, \emph{Phys. Rev. B}, {\bf 68}, 125415 (2003).
\bibitem{PhysRevLett.90.040401} Bulaevskii, L.N. \& Ortiz, G., Tunneling Measurement of a Single Quantum Spin, \emph{Phys. Rev. Lett.} {\bf 90}, 040401 (2003).
\bibitem{manassen2004} Manassen, Y. \& Balatsky, A.V., 1/f Spin Noise and a Single Spin Detection with STM, \emph{Isr. J. Chem.} {\bf 44}, 401 (2004).
\bibitem{balatsky2006} Balatsky, A.V., Fransson, J., Mozyrsky, D., \& Manassen, Y., STM NMR and nuclear spin noise, \emph{Phys. Rev. B}, {\bf 73}, 184429 (2006).
\bibitem{golub2013} Golub, A. \& Horovitz, B., Nanoscopic interferometer model for spin resonance in current noise, \emph{Phys. Rev. B}, {\bf 88}, 115423 (2013).

\bibitem{Science.350.417} Baumann, S. et al.
Electron paramagnetic resonance of individual atoms on a surface, \emph{Science}, {\bf 350}, 417--420 (2015).

\bibitem{billangeon2007} Billangeon, P.-M., Pierre, F., Bouchiat, H., \& Deblock, R., Very High Frequency Spectroscopy and Tuning of a Single-Cooper-Pair Transistor with an On-Chip Generator, \emph{Phys. Rev. Lett.} {\bf 98}, 126802 (2007).
\bibitem{bretheau2013} Bretheau, L., Girit, \c{C}. \"O ., Pothier, H., Esteve, D., \& Urbina, C., Exciting Andreev pairs in a superconducting atomic contact, \emph{Nature}, {\bf 499} 312 (2013).

\bibitem{chiesa2013} Chiesa, A., Carretta, S., Santini, P., Amoretti, G., \& Pavarini, E., Many-Body Models for Molecular Nanomagnets, \emph{Phys. Rev. Lett.} {\bf 110}, 157204 (2013).
\bibitem{coronado2004} Coronado, E. \& Day, P., Magnetic Molecular Conductors, \emph{Chem. Rev.} {\bf 104}, 5419--5448 (2004).
\bibitem{chen2008} Chen, X. et al.
Probing Superexchange Interaction in Molecular Magnets by Spin-Flip Spectroscopy and Microscopy, \emph{Phys. Rev. Lett.} 101, 197208 (2008).
\bibitem{mugarza2012} Mugarza, A. et al.
Electronic and magnetic properties of molecule-metal interfaces: Transition-metal phthalocyanines adsorbed on Ag(100), \emph{Phys. Rev. B}, {\bf 85}, 155437 (2012).
\bibitem{krull2013} Krull, C., Robles, R., Mugarza, A., \& Gambardella, P., Site- and orbital-dependent charge donation and spin manipulation in electron-doped metal phthalocyanines, \emph{Nature Mater.} {\bf 12}, 337 (2013).

\bibitem{NanoLett.9.2414} Fransson, J., Spin Inelastic Electron Tunneling Spectroscopy on Local Spin Adsorbed on Surface, \emph{Nano Lett.} {\bf 9}, 2414--2417 (2009).
\bibitem{PhysRevB.81.115454} Fransson, J., Eriksson, O., \& Balatsky, A.V., Theory of spin-polarized scanning tunneling microscopy applied to local spins, \emph{Phys. Rev. B}, {\bf 81}, 115454 (2010).

\bibitem{PhysRevLett.102.256802} Fern\'andez-Rossier, J., Theory of Single-Spin Inelastic Tunneling Spectroscopy, \emph{Phys. Rev. Lett.} {\bf 102} 256802 (2009).
\bibitem{PhysRevLett.103.050801} Persson, M., Theory of Inelastic Electron Tunneling from a Localized Spin in the Impulsive Approximation, \emph{Phys. Rev. Lett.} {\bf 103}, 050801 (2009).
\bibitem{PhysRevLett.103.176601} Lorente, N. \&Gauyacq, J.-P. , Efficient Spin Transitions in Inelastic Electron Tunneling Spectroscopy, \emph{Phys. Rev. Lett.} {\bf 103}, 176601 (2009).

\bibitem{NatPhys.9.765} Heinrich, B. W., Braun, L., Pascual, J. I., \& Franke, K. J., Protection of excited spin states by a superconducting energy gap, \emph{Nature Phys.} {\bf 9}, 765 (2013).
\bibitem{NanoLett.15.4024} Heinrich, B. W., Braun, L., Pascual, J. I., \& Franke, K. J., Tuning the Magnetic Anisotropy of Single Molecules, \emph{Nano Lett.} {\bf 15}, 4024 (2015).

\bibitem{PhysRevLett.92.107001} Zhu, J.-X., Nussinov, Z., Shnirman, A., \& Balatsky, A. V., Novel Spin Dynamics in a Josephson Junction, \emph{Phys. Rev. Lett.} {\bf 92}, 107001 (2004).
\bibitem{NewJPhys.10.013017} Fransson, J. \& Zhu, J.-X., Spin dynamics in a tunnel junction between ferromagnets, \emph{New J. Phys.} {\bf 10}, 013017 (2008).
\bibitem{PhysRevLett.108.057204} Bhattacharjee, S., Nordstr\"om, L., \& Fransson, J., Atomistic Spin Dynamic Method with both Damping and Moment of Inertia Effects Included from First Principles, \emph{Phys. Rev. Lett.} {\bf 108}, 057204 (2012).
\bibitem{PhysRevLett.113.257201} Fransson, J., Ren, J., \& Zhu, J.-X., Electrical and Thermal Control of Magnetic Exchange Interactions, \emph{Phys. Rev. Lett.} {\bf 113}, 257201 (2014).

\bibitem{PhysRevB.92.125405} Fransson, J., Inelastic-impurity-scattering-induced spin texture and topological transitions in surface electron waves, \emph{Phys. Rev. B}, {\bf 92}, 125405 (2015).

\bibitem{PhysRevLett.115.237202} Baumann, S. et al.
Origin of Perpendicular Magnetic Anisotropy and Large Orbital Moment in Fe Atoms on MgO, \emph{Phys. Rev. Lett.} {\bf 115}, 237202 (2015).
\bibitem{Science.344.988} Rau, I. G. et al.
Reaching the magnetic anisotropy limit of a 3d metal atom, \emph{Science}, {\bf 344}, 988--992 (2014).

\bibitem{EPL.108.67009} Berggren, P. \& Fransson, J., Spin inelastic electron tunneling spectroscopy on local magnetic moment embedded in Josephson junction, \emph{EPL}, {\bf 108}, 67009 (2014).
\bibitem{PhysRevB.91.205438} Berggren, P. \& Fransson, J., Theory of spin inelastic tunneling spectroscopy for superconductor-superconductor and
superconductor-metal junctions, \emph{Phys. Rev. B}, {\bf 91}, 205438 (2015).

\end{thebibliography}





\end{document}